\documentstyle[referee,psfig,epsf]{l-aa}
\def\lsim{\lower.5ex\hbox{$\; \buildrel < \over \sim \;$}}
\def\gsim{\lower.5ex\hbox{$\; \buildrel > \over \sim \;$}}
\newcommand{\eqb}{\begin{eqnarray}}
\newcommand{\eqe}{\end{eqnarray}}

\begin{document}
\thesaurus{02.01.2, 02.02.1, 08.14.1, 02.08.1}
\title{Bondi Flows on Compact Objects Revisited}
\author{Sandip K. Chakrabarti\inst{1}, Soumitri A. Sahu\inst{2}}
\institute{SNBNCBS, Calcutta 700091 and TIFR, Mumbai 400005 INDIA
\and
University of Hyderabad, Gachibowli, Hyderabad  500046 INDIA}

\offprints{S. K. Chakrabarti, S.N. Bose National Center for Basic Sciences, JD Block, Salt Lake, Calcutta 700091,
INDIA}
\date{Received Oct 15th, 1996 ; accepted Dec. 10th, 1997}
\maketitle
\markboth{Bondi Flows Revisited}{}

%\begin{document}

\maketitle

\begin{abstract}

We revisit Bondi flows on black holes, naked singularities and neutron stars
keeping in mind recent progress in understanding of the flow behaviour near 
compact objects. In an {\it adiabatic spherical accretion} on black holes 
and naked singularities, standing shock waves do not form, but shocks on 
neutron stars may or may not form depending on the boundary conditions. 
For low accretion rates (hard state), these objects may behave similarly.
However, for high accretion rates (soft state), the converging flow
on a black hole will produce a weak hard tail of energy spectral slope
of $\sim 1.5$ (apart from the soft bump) while the flow on a neutron star or 
a naked singularity will have no hard tail. The soft bumps of neutron star 
candidates may generally be at higher energies than those of naked singularities. 

\keywords
{accretion, accretion disks
--- black hole physics
--- stars: neutron 
--- hydrodynamics}
\end{abstract}

\section{Introduction}

Spherical flows have been studied in the past four decades 
rather extensively (Bondi, 1952; Shapiro, 1973; Shapiro \& Salpeter, 1975;
Park \& Ostriker, 1989; see Chakrabarti, 1996a for a recent review).
These zero angular momentum flows were not found to be very
efficient radiators since they carry the total energy
along with them, except at the boundary layer of
objects with hard surfaces. Therefore, it was difficult to
explain, for example, the quasar luminosity and the soft states of 
galactic and extra-galactic black hole candidates using spherical
flow models around black holes. The Keplerian accretion disks (Shakura \& Sunyaev, 1973),
on the other hand, are very efficient radiators. They locally dissipate
the entire amount of heat that is generated by viscosity.
Generalized adiabatic Bondi flows which contained
angular momentum (Liang \& Thompson, 1980; Chakrabarti, 1989; Chakrabarti
1990a, hereafter C90a) could also be highly inefficient emitters. However, such flows 
with viscosity and cooling effects which may join a Keplerian 
or sub-Keplerian flow far away can have intermediate efficiencies 
(C90a; Chakrabarti, 1990b hereafter C90b; Chakrabarti 1996ab, hereafter C96ab). 
In a black hole accretion, they partly carry along energy through the event
horizon (see, Fig. 8a of C90a) and partly radiate the dissipated
heat and thus successfully bridge the gap between a classical Bondi
flow and a classical Shakura-Sunyaev type Keplerian disk. Shocks
may also form just outside the horizon if the flow can pass through
two sonic points. In a neutron star accretion, the flow dissipates 
the energy carried along in the boundary layer just outside the neutron 
star surface (Chakrabarti 1989; C96b). These disks 
which may contain both Keplerian and sub-Keplerian flows can 
explain most of the stationary and non-stationary spectral features 
of black hole and neutron star candidates as explained in detail in
Chakrabarti \& Titarchuk (1995, hereafter CT95); C96b; Molteni et al.
(1996); Ryu et al. (1997).

In the present paper, we take a closer look at spherical accretion
flows, not only on black holes (BH), but also on other compact
objects, such as weakly magnetized and slowly rotating neutron stars (NS)
and naked singularities (NSing).
Naked singularities are not as widely considered to be
astrophysically relevant as black holes and neutron stars, 
though recently their theoretical existence have been discussed, 
and serious studies of using them to explain astrophysical 
phenomena are being considered (Penrose, 1974; Ori \& Piran 1990; 
Nakamura et al. 1993; Chakrabarti \& Joshi, 1994).
The adiabatic accretion flow onto black holes and naked singularities
has to pass through a sonic surface and enters the horizon or the 
singularity supersonically, while on an unmagnetized neutron stars, the flow may 
or may not have standing shock waves as it rapidly lands on the
surface. We discuss the fundamental differences between these solutions 
and present some possible spectral signatures by which these objects 
could be distinguished. This review of the subject was essential
in view of the recognition that the quasi-spherical sub-Keplerian flows could
be useful in explaining the quiescent states of black holes
(Ebisawa et al, 1996), hard states of black holes and neutron
stars and more importantly weak hard tails seen in the soft states 
of black hole candidates (CT95).

In the next Section, we present the basic equations, and
the procedure to solve them. In \S 3, we present the
solutions of these equations. In \S 4, we briefly discuss the
spectral properties of these solutions. Finally, in \S 5, we
discuss importance of angular momentum and its effect
on the results and make concluding remarks.

\section{Basic Equations}

We simplify our hydrodynamic calculations around a black hole and 
a neutron star by choosing the Paczy\'nski-Wiita (1980) pseudo-potential,
$$
\phi = - {1\over{(x-2)}}
\eqno{(1a)}
$$
which mimics the geometry around a compact star quite well.
Around the naked singularity we choose an ordinary Newtonian potential
$$
\phi =- {1\over x}    ,
\eqno{(1b)}
$$
allowing matter to flow arbitrarily close to $x=0$. The main conclusions
drawn in this paper are not affected by these simplifying assumptions.
However, intuitively, we do assume that the inner boundary condition 
on accretion flows into a naked singularity is supersonic, which is valid 
if the flow does not violate causality (velocity of sound less than the
velocity of light). 

In an adiabatic Bondi flow, the conserved specific energy is given by (C90b)
$$
{\cal E} = {1\over 2} u^2  +  n a^2  -  {(1-C)\over(x - x_0)}  = n {a_\infty}^2
\eqno{(2)}
$$
using $G=M_{BH}=c=1$, where $G$ is the gravitational constant,
$M_{BH}$ is the mass of the compact object, and $c$ is the velocity of light,
so that the units of mass, length and time are $M_{BH}$, $GM_{BH}\over{c^2}$ 
and $GM_{BH}\over{c^3}$ respectively. Here $u$ is the radial velocity,
$x$ is the spherical radial coordinate, $a$ is the adiabatic sound speed,
$a^2=\gamma P/\rho $ and  $\gamma$ is the adiabatic index, $P$ is the pressure,
$\rho$ is the mass density, $n=(\gamma-1)^{-1}$, and $x_0$ is either $2$
or $0$ depending on the nature of the compact object. Equation (2) is obtained
by integrating the radial momentum equation. Note that we
have weakened the gravitational potential by $C/(x-x_0)$ which 
essentially represents the outward radiative force $C/(x-x_0)^2$.
Here, $C$ is chosen to be independent of $x$, for convenience. Roughly
speaking, on a neutron star accretion, $C$ scales with accretion rate:
$C\sim {\dot M}/{\dot M}_{Edd}$. In a black hole or naked singularity
accretion, radiation could be partly trapped and drawn in radially, and therefore
there is no simple relation between $C$ and the accretion rate. On the right
hand side we have $a_\infty$ which is the sound speed 
at a large distance. If the flow originates from a Keplerian
flow (C90ab, C96ab), it may have to be pre-heated (either by radiation
or by magnetic dissipation) to make ${\cal E}>0$,
in order that it passes through a sonic point at a large distance.

\noindent The mass flux is obtained as,
$$
\dot M = 4 \pi \rho u x^2,
\eqno{(3a)}
$$
which could be re-written in terms of `entropy-accretion' flux
$$
{\dot {\cal M}}= {\dot M} (\gamma K)^n/ 4\pi= a^{2n} u x^2.
\eqno{(3b)}
$$
Here, $K=P/\rho^\gamma$ is a constant and a measure of
entropy and therefore $K$ can change only if the flow
passes through a shock. Thus, ${\dot {\cal M}}$ is a 
measure of entropy and mass flux. From Eqs. 2 and 3, 
$$
{du\over{dx}} = {{{{(1-C)}\over(x - x_0)^2} - {2a^2\over x}}\over
{{a^2\over u} - u}}.
\eqno{(4)}
$$
At the {\it sonic surface}, where numerator and denominator vanish, 
one must have,
$$
u_c = a_c,
\eqno{(5a)}
$$
and,
$$
x_c = x_0 + {{(1-C)}\over{4 a_c^2}} + [{{x_0(1-C)}\over{2 a_c^2}}+
{{(1-C)^2}\over{16 a_c^4}}]^{1/2} .
\eqno{(5b)}
$$
We have ignored the other positive root for $x_c$, as it is nearly zero.
Flow can be sonic at $x_c=0$ for a naked singularity ($x_o=0$)
and therefore is inside the horizon or the star surface.

Since two conditions (5a) and (5b) are to be satisfied, while
only one extra unknown (namely, $x_c$) is introduced, clearly,
both the specific energy ${\cal E}$ and entropy-accretion flux 
${\dot {\cal M}}$ cannot be independent. Indeed, for a given energy ${\cal E}$,
the critical entropy-accretion flux is,
$$
{\dot {\cal M}_{c}}=[\frac{2}{3}({\cal E} + 
\frac{1-C}{x_c-x_0})]^{n+1/2} x_c^2 .
\eqno{(6)}
$$

It is clear that if the flow is {\it hot} at a large
distance ($a_\infty \gsim 0$), i.e., ${\cal E}\gsim 0$,
then the sonic surface $x=x_c$ is located at a finite distance
provided the polytropic index is suitable (roughly,
$\gamma <5/3$. See, C90b).
Once the flow passes through the sonic surface it will 
continue to remain supersonic if the central object is a black 
hole or a naked singularity, but the flow has to pass through a 
standing shock (at $x=x_s$, say) and become subsonic if the 
central object is a neutron star. On the other hand,
a neutron star accretion can be subsonic throughout if 
$$
{\dot {\cal M}} <{\dot {\cal M}_c}.
\eqno{(7a)}
$$
If on the surface of the neutron star,
$$
{\dot {\cal M}}|_{r_*} >{\dot{\cal M}_c}
\eqno{(7b)}
$$
then the flow must have a standing shock and the entropy
generated at the shock must be such that the post-shock flow has
$$
{\dot {\cal M}}|_{r_s}= {\dot {\cal M}}|_{r_*} .
\eqno{(8)}
$$
This condition along with the pressure balance condition
$$
P_-+\rho_- u_-^2 =P_++\rho_+ u_+^2
\eqno{(9)}
$$
determines the location of the standing shock.

The Eqs. 7a, 7b and 8 can be translated 
in many ways in terms of the injection speed, temperature, the 
gradient of velocities at the outer or inner boundary, or the 
location of the sonic point by employing any of the
definitions given by Eq. 2, Eq. 3b, Eq. 4 or Eq. 6.

\section {Solutions of the Basic Equations}

\begin{figure}
\vbox{
\vskip -2.0cm
\centerline{
\psfig{figure=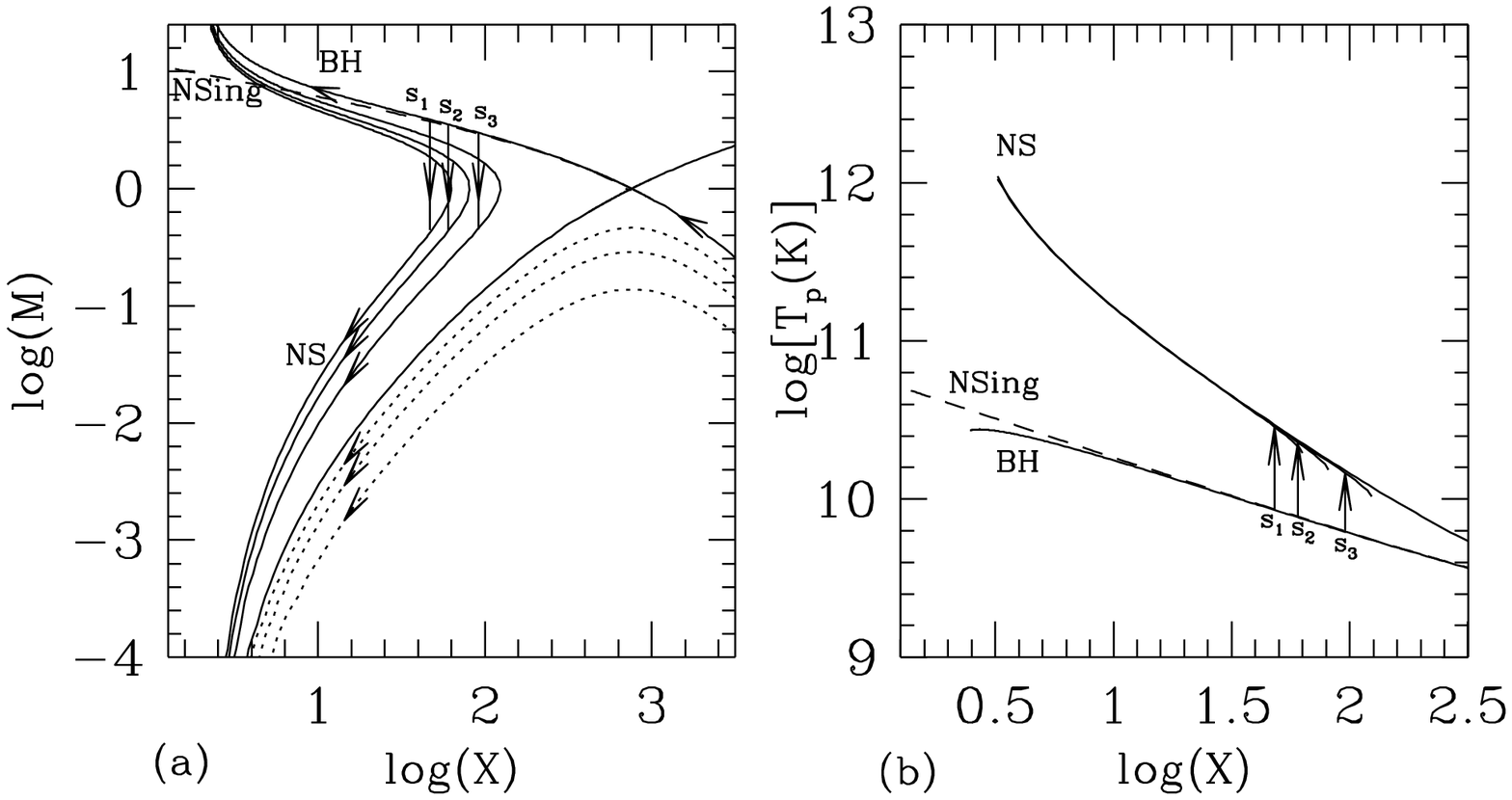,height=15.0truecm,width=15.0truecm,angle=0}}}
\vspace {-4.0cm}
\noindent {\small {\bf Fig. 1ab:}
(a) Mach number variation with radial distance. Vertical arrows are shock
locations in the neutron star accretion. Different arrows are for 
different entropy accretion rates on neutron stars. Solutions on black holes
(BH, solid), naked singularities (NSing, long dashed), and neutron stars (NS,
dotted and subsonic solid) are distinguished. (b) Proton temperatures
in corresponding solutions. Subsonic neutron star accretion flow is much hotter.}
\end{figure}
\begin{figure}
\vbox{
\vskip -2.0cm
\centerline{
\psfig{figure=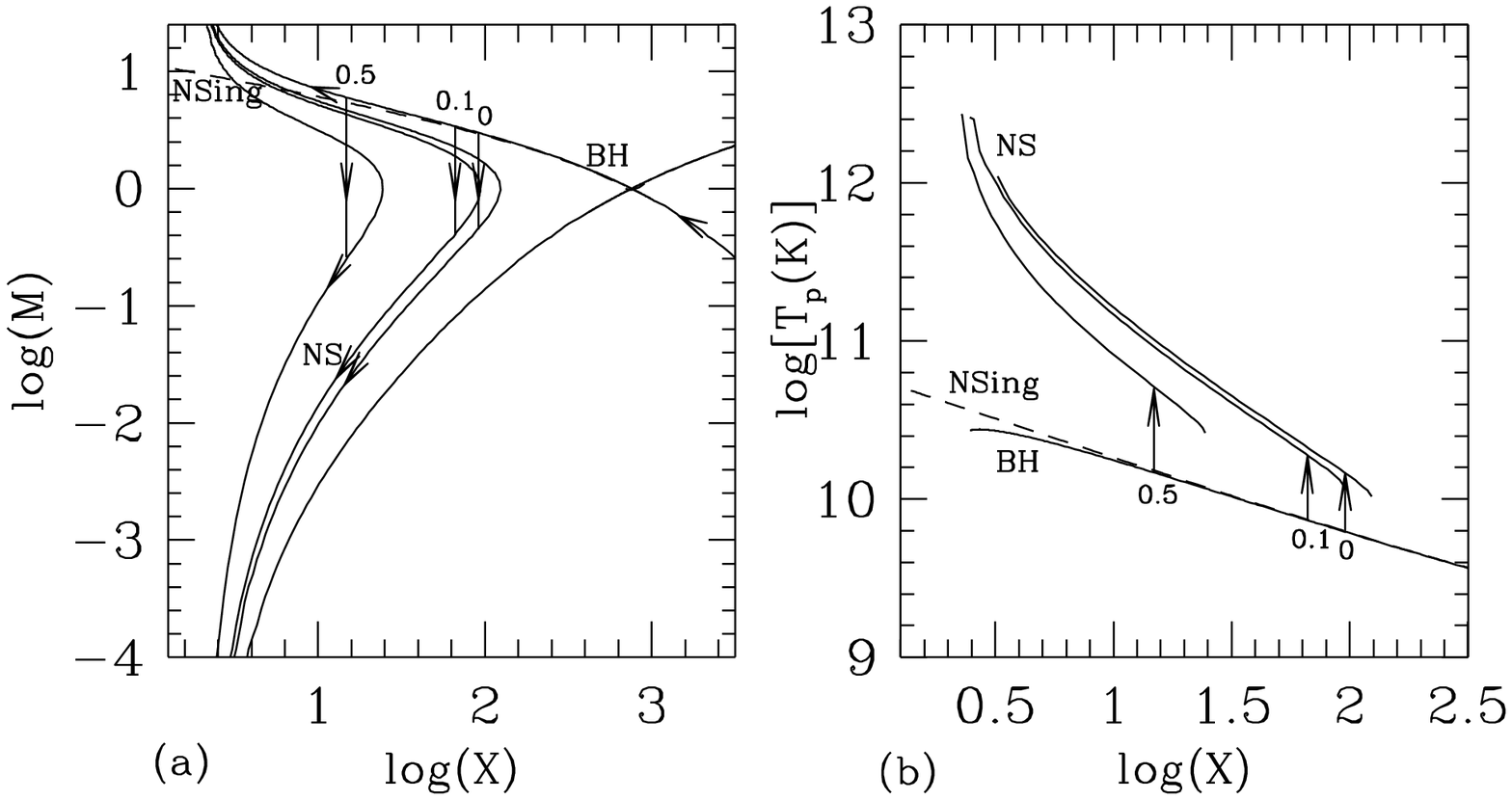,height=15.0truecm,width=15.0truecm,angle=0}}}
\vspace {-4.0cm}
\noindent {\small {\bf Fig. 2ab:}
(a) Similar to Fig. 1ab, except that the radiation pressure
(parametrized by $C=0,\ 0.1,\ 0.5$ marked on curves)
on the neutron star surface is varied.
(b) Proton temperatures in corresponding solutions.}
\end{figure}
We now present the complete set of solutions obtained using fourth order
Runge-Kutta method. Fig. 1a shows the variation of the
Mach number with radial distance (in logarithmic coordinates 
in both directions). The solid curves intersecting at the sonic 
point $x_c$ are the well known Bondi solutions. This has ${\dot{\cal M}_c}=
4.315 \times 10^{-6}$. The arrowed branch which becomes supersonic on the horizon
represents the black hole accretion. The long dashed curve
(marked NSing) is drawn with $x_0=0$ (with same ${\dot {\cal M}_c}$) and 
represents an accretion on a naked singularity. The accretion flows 
on neutron star surfaces must be subsonic. The exact subsonic branch
depends on the inner boundary condition, i.e., the way matter lands 
on the staller surface (say, the derivative of the velocity on the 
surface.). The solid arrowed curves leading to the neutron star surface 
have more entropy than the curve passing through the critical point 
(${\dot {\cal M}}|_{r_*}>{\dot{\cal M}}_c$). The difference in entropy
${\dot {\cal M}}|_{r_*} -{\dot{\cal M}}_c$ must be generated
at the standing shocks located at $s_1$, $s_2$ and $s_3$
respectively. In the present example, ${\dot{\cal M}}_{r_*}=
3.4315 \times 10^{-5}$, $2.4315 \times 10^{-5}$ and $1.4315 \times 10^{-5}$
(for the solid curves drawn in the order inside to outside). The corresponding
shock locations are $s_1= 46.77$, $s_2=60.0$ and $s_3=91.6$ respectively. Here,
$\gamma=4/3$ (relativistic flow) and $C=0$ (no excess radiation pressure) is chosen. 
These correspond to low accretion rate solutions.
The short dashed arrowed curves leading to
a neutron star have even less entropy than ${\dot{\cal M}}_c$. In this
case, they are drawn for ${\dot{\cal M}}_{r_*}= 10^{-6}, \ 2\times 10^{-6}$ 
and $ 3\times 10^{-6}$ (from bottom to top) respectively. In Fig. 1b,
the proton temperatures $T_p$ (in degrees Kelvin) for the
corresponding solutions are plotted
on a logarithmic scale as a function of the logarithmic radial
distance. The black hole and naked singularity solutions
have lower temperatures since they are supersonic, while
the adiabatic subsonic neutron star solutions have higher 
temperature, and the distributions are almost independent 
of the branch that is chosen.
As the entropy of the post-shock flow is increased,
the shock location comes closer to the neutron star surface
and the post-shock temperature is also increased.
\begin{figure}
\vbox{
\vskip -2.0cm
\centerline{
\psfig{figure=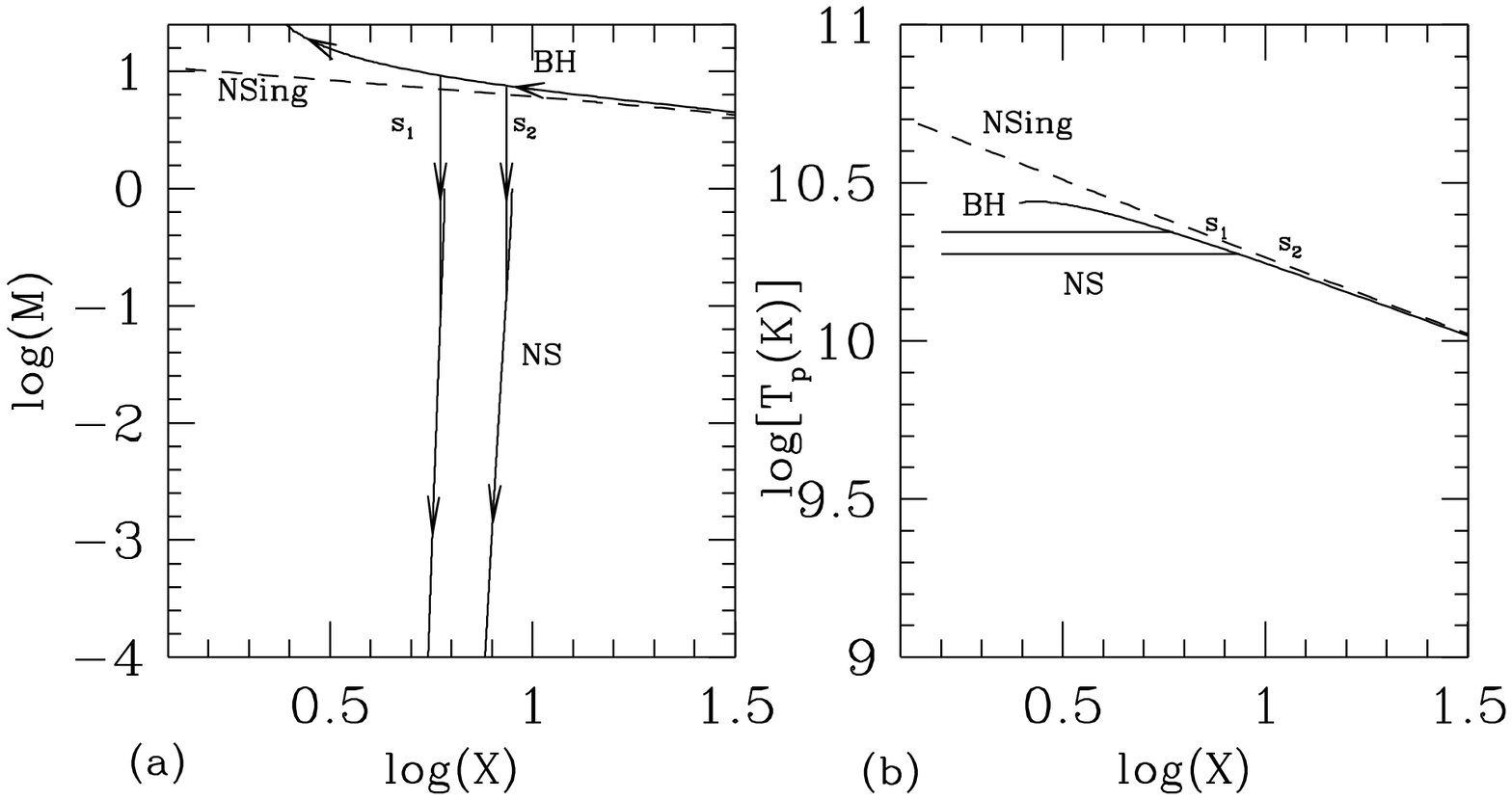,height=15.0truecm,width=15.0truecm,angle=0}}}
\vspace {-4.0cm}
\noindent {\small {\bf Fig. 3ab:}
(a) Similar to Fig. 1ab, except that the post-shock
region is cooler and isothermal due to dissipation.
(b) Proton temperatures in corresponding solutions.}
\end{figure}
\begin{figure}
\vbox{
\vskip -2.0cm
\centerline{
\psfig{figure=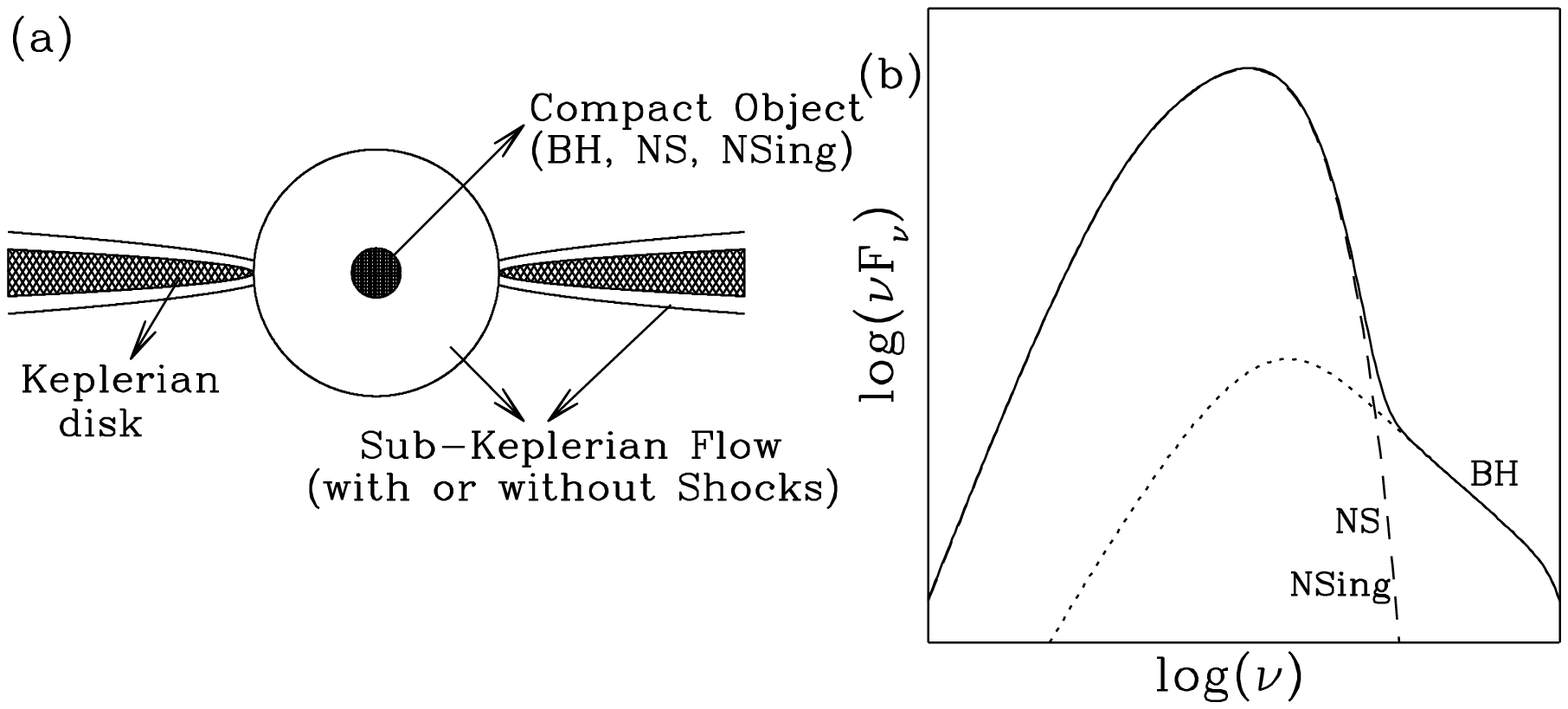,height=15.0truecm,width=12.0truecm,angle=0}}}
\vspace {-4.0cm}
\noindent {\small {\bf Fig. 4ab:}
(a) Schematic view of generalized solutions on compact objects where flows close to
the objects are sub-Keplerian Bondi type (with or without shocks).
(b) Variation of the soft state spectra in three types of compact objects.
In neutron stars (NS) and naked singularities (NSing) the soft state spectra should
not have the weak hard tail, while the black holes should have this 
feature.}
\end{figure}
In Figs. 2(a-b), we include the effect of the radiative force
to study the flow properties of the solutions on neutron stars. 
In Fig. 2a, we plot three solid curves with arrows leading to the 
neutron star surface. 
We choose ${\dot{\cal M}}_{r_*}=1.4315 \times 10^{-5}$ 
as the post-shock entropy accretion rate on the surface.
The vertical arrows are drawn at shock locations
and at each shock, the value of $C$ (such as $0$, $0.1$, and $0.5$)
is marked. The value of $C$ is made different from zero
only in the post-shock region, since the subsonic
flow would be maximally affected by the radiation pressure effects.
The corresponding shock locations are $91.54$ (same as in
Fig. 1a), $66.1$ and $14.8$ respectively.
In Fig. 2b, temperature distributions are plotted. Note that 
as $C$ is increased, the temperature of post-shock region
becomes smaller. This is because when the radiation pressure is 
present, one does not require thermal pressure very much. In the pre-shock
region, the ram pressure must increase sufficiently to balance the
net pressure. As a result, the shock is located at a place closer 
to the neutron star surface.

In Fig. 3(a-b), we consider the case when the post-shock region is 
isothermal. In this case, the shock will be in pressure
equilibrium only if the neutron star surface is much cooler
than the cases mentioned above. The surface temperature of the neutron 
star (which is also the post-shock temperature of the flow) determines the
shock location. Two horizontal lines have sound speeds $a=0.06$ (lower) and
$a=0.065$ respectively. When the pressure balance condition is satisfied
at the shocks, these sound speeds determine the post-shock Mach number variation
which go to zero very rapidly close to the surface. As the
surface temperature of the star is lowered, the location of the
shock, i.e, the width of the terminal boundary layer is also increased
since the pre-shock flow  temperature is monotonic. 

\section {Spectral Signatures of Compact Objects}

In Fig. 4a, we schematically show our present understanding of
the accretion flows onto compact objects. The Keplerian
or sub-Keplerian flows at the outer boundary become
increasingly sub-Keplerian close to the inner boundary. On the black hole
horizon, and presumably at the naked singularity,
the radial velocity approaches the velocity of light
(C90b; Chakrabarti 1996c) and since  even in the
extreme case the sound speed is less than this, the
flow must be supersonic at the inner boundary.
In the case of accretion onto a neutron star however, the
flow is subsonic on the inner boundary. 
CT95 and Titarchuk et al. (1996) pointed out that in the limit of high
accretion rates (where the Thompson scattering opacity
$\tau \gsim 1$) the eigenvalue of the corresponding
radiative transfer equation in presence of convergent fluid
flow is given by,
$$
\lambda^2 = \frac{3}{2} + \frac{3}{4} \frac {1-C}{1+C} X_b^{1/2}
$$
where, $X_b$ is the location of the inner boundary in units 
of the Schwarzschild radius. In terms of this eigen value, 
the energy spectral index $\alpha$ [$F(\nu) \sim \nu^{-\alpha}$] 
is found out to be,
$$
\alpha = 2 \lambda^2 - 3 .
$$

In the case of a black hole accretion: $X_b=1$ and $C=0$; and
one obtains $\alpha=1.5$, very similar to what is observed
in black hole candidates (Sunyaev et al. 1994; Ebisawa et al, 1996).
If $X_b=1.5$ is chosen (which corresponds to the last photon orbit
and is probably more physical) the corresponding value is $\alpha\sim 1.83$.
In the case of high accretion rates on neutron stars,
$C \sim 1$ giving rise to $\alpha=0$. The spectrum at higher energies is 
thus completely flat. In the case of a naked singularity, $X_b=0$, 
and the spectrum in higher energies is also flat, 
independent of the accretion rate as long as the 
post-Keplerian flow (with or without shock waves) 
has $\tau \gsim 1$. Naked singularities may be similar 
to black holes in the hard states, when the accretion rates 
are smaller, while similar to neutron stars in soft states 
although the soft bump is cooler unless the neutron star 
accretion itself is isothermal (as is possible 
when shocks are present; see, Shapiro \& Salpeter, 1975). These 
aspects of spectral properties are shown qualitatively in Fig. 4b.

As far as the total energy of radiation is concerned, one could distinguish 
these objects as well. Since the flow has to dissipate its energy 
at the hard surface of a neutron star, the luminosity
of neutron stars would still be proportional to the accretion rates
provided magnetic field is weak enough; non-linear 
interaction with magnetic fields (Illarionov, \& Sunyaev, 1975) may 
change this conclusion. On the contrary, in a black hole accretion, 
luminosity could be very small since the flow disappears through the horizon
(for instance, constant energy flows of C89 have, strictly speaking, zero luminosity). 
In a naked singularity, there is no horizon, and extremely hot matter
very close to the origin may cause thermonuclear flashes and matter
should be at least partially luminous also. However, the detailed physics is 
not very well understood since one must take into account quantum effects.

\section {Concluding Remarks}

Our understanding of the accretion processes on compact objects
is still far from complete. In the present paper, we demonstrated
that the outcome of solutions of spherical hydrodynamic accretion
processes in terms of spectral properties can differ
very much depending upon the behavior of matter
at the inner boundary. We have discussed the possibility that the
weak hard tail in the soft states of a black hole is
a result of the quasi-spherical flows and the hard tail
should be absent in neutron stars and naked singularities.
We also showed (Fig. 1b) that  on account of their
subsonic inner boundary condition, neutron stars produce much hotter radiation 
that the black holes or naked singularities. It is interesting to note that
although the spherical flows are radiatively inefficient, they
definitely have to dissipate their energy at the boundary layer,
namely, in the post-shock region if the central object is a neutron
star. The quasi-periodic oscillation from neutron stars
could very well be due to the oscillation of this post-shock 
region (Molteni et al 1996). This is true even when some magnetic field is present
on the neutron star surface as the radial flows may penetrate fields easier. 
For a black hole accretion such restriction on efficiency is not present
as the flow can carry along most of its energy through the horizon.

Although our study is strictly applicable only for spherical flows, 
should some angular momentum be present, the solution would look similar to what 
is shown in Fig. 4a, which is a combination of Keplerian and 
sub-Keplerian flows (with or without shocks) as discussed in 
Chakrabarti \& Titarchuk (1995) and Chakrabarti (1996b). 
Close to the inner boundary, the circular orbits are 
unstable anyway, and thus the flow would become quasi-spherical
even in presence of rotation. 

{\sl Acknowledgments:}

SS acknowledges the hospitality of Tata Institute VSRP programme
in the summer of 1996 during which this work was carried out. 

{}

\end{document}